\documentclass[aps,pre,showpacs,twocolumn,groupedaddress,floatfix]{revtex4}
\usepackage{amssymb,amsfonts}
\usepackage[intlimits]{amsmath}
\usepackage{graphics,epsfig}

\def\ni{\noindent}
\def\etal{et al}

\def\Svd{\Ex{\ov{v\dotv}}}
\def\Tvdx{\Ex{v\dotv}_x}
\def\Mvdx{\Ex{\wh{v\dotv}}_x}
\def\Mvd{\Ex{\wh{v\dotv}}}

\def\kv{k_v}

\def\ks{k_\ms}
\def\qv{\hat{k}_v}

\def\qs{\hat{k}_\ms}

\def\lamv{\lambda_v}
\def\lams{\lambda_\ms}
\def\lame{\lambda_\me}
\def\lami{\lambda_\mi}
\def\laml{\lambda_\ell}

\def\Mvhsx{\Ex{\wh{vh_\ms}}_x}
\def\Mvhex{\Ex{\wh{vh_\me}}_x}
\def\Mvhix{\Ex{\wh{vh_\mi}}_x}
\def\Mvhs{\Ex{\wh{vh_\ms}}}
\def\Mvhe{\Ex{\wh{vh_\me}}}
\def\Mvhi{\Ex{\wh{vh_\mi}}}

\def\Thhsx{\Ex{h^2_\ms}_x}

\def\Shhsx{\Ex{\ov{h^2_\ms}}_x}

\def\Shhe{\Ex{\ov{h^2_\me}}}
\def\Shhi{\Ex{\ov{h^2_\mi}}}
\def\Shhs{\Ex{\ov{h^2_\ms}}}
\def\Mhhsx{\Ex{\wh{h^2_\ms}}_x}
\def\Mhhex{\Ex{\wh{h^2_\me}}_x}
\def\Mhhix{\Ex{\wh{h^2_\mi}}_x}
\def\Mhhe{\Ex{\wh{h^2_\me}}}
\def\Mhhi{\Ex{\wh{h^2_\mi}}}

\def\Tvhex{\Ex{vh_\me}_x}
\def\Tvhix{\Ex{vh_\mi}_x}
\def\Tvhsx{\Ex{vh_\ms}_x}
\def\Svhsx{\Ex{\ov{vh_\ms}}_x}
\def\Svhex{\Ex{\ov{vh_\me}}_x}
\def\Svhix{\Ex{\ov{vh_\mi}}_x}
\def\Svhs{\Ex{\ov{vh_\ms}}}
\def\Svhe{\Ex{\ov{vh_\me}}}
\def\Svhi{\Ex{\ov{vh_\mi}}}
\def\Mvhsx{\Ex{\wh{vh_\ms}}_x}
\def\Mvhex{\Ex{\wh{vh_\me}}_x}
\def\Mvhix{\Ex{\wh{vh_\mi}}_x}
\def\Mvhs{\Ex{\wh{vh_\ms}}}
\def\Mvhe{\Ex{\wh{vh_\me}}}
\def\Mvhi{\Ex{\wh{vh_\mi}}}

\def\Tvhex{\Ex{vh_\me}_x}
\def\Tvhix{\Ex{vh_\mi}_x}
\def\Tvhsx{\Ex{vh_\ms}_x}
\def\Svhsx{\Ex{\ov{vh_\ms}}_x}
\def\Svhex{\Ex{\ov{vh_\me}}_x}
\def\Svhix{\Ex{\ov{vh_\mi}}_x}
\def\Svhs{\Ex{\ov{vh_\ms}}}
\def\Svhe{\Ex{\ov{vh_\me}}}
\def\Svhi{\Ex{\ov{vh_\mi}}}
\def\Mvhsx{\Ex{\wh{vh_\ms}}_x}
\def\Mvhex{\Ex{\wh{vh_\me}}_x}
\def\Mvhix{\Ex{\wh{vh_\mi}}_x}
\def\Mvhs{\Ex{\wh{vh_\ms}}}
\def\Mvhe{\Ex{\wh{vh_\me}}}
\def\Mvhi{\Ex{\wh{vh_\mi}}}

\def\Sdhex{\Ex{\ov{\dotv h_\me}}_x}
\def\Sdhix{\Ex{\ov{\dotv h_\mi}}_x}
\def\Sdhe{\Ex{\ov{\dotv h_\me}}}
\def\Sdhi{\Ex{\ov{\dotv h_\mi}}}
\def\Sdhs{\Ex{\ov{\dotv h_\ms}}}

\def\Mdhex{\Ex{\wh{\dotv h_\me}}_x}
\def\Mdhix{\Ex{\wh{\dotv h_\mi}}_x}
\def\Mdhe{\Ex{\wh{\dotv h_\me}}}
\def\Mdhi{\Ex{\wh{\dotv h_\mi}}}

\def\Tvvx{\Ex{v^2}_x}
\def\Svvx{\Ex{\ov{v^2}}_x}
\def\Svv{\Ex{\ov{v^2}}}
\def\Svvsx{\Ex{\ov{v^2}}_x^\ms}
\def\Svvex{\Ex{\ov{v^2}}_x^\me}
\def\Svvix{\Ex{\ov{v^2}}_x^\mi}
\def\Mvvx{\Ex{\wh{v^2}}_x}
\def\Mvvsx{\Ex{\wh{v^2}}_x^\ms}

\def\Mvv{\Ex{\wh{v^2}}}

\def\Mvve{\Ex{\wh{v^2}}^\me}
\def\Mvvi{\Ex{\wh{v^2}}^\mi}

\def\MEs{ \wh{{{\cal E}}}_\mathrm{s}}
\def\MEe{ \wh{{{\cal E}}}_\mathrm{e}}
\def\MEi{ \wh{{{\cal E}}}_\mathrm{i}}

\def\Ex#1{\langle#1\rangle}

\def\l{\left}
\def\r{\right}

\def\be{\begin{eqnarray}}
\def\ee{\end{eqnarray}} 

\def\p{\partial}

\def\w{\omega}

\def\ov#1{\overline{#1}}

\def\wh#1{\widehat{#1}}

\def\erf{\mathrm{erf}}

\def\me{\mathrm{e}}
\def\mi{\mathrm{i}}
\def\ml{\mathrm{\ell}}
\def\ms{\mathrm{s}}
\def\mth{\mathrm{th}}

\def\tauv{\tau_v}
\def\taus{\tau_\ms}
\def\taue{\tau_\me}
\def\taui{\tau_\mi}
\def\taul{\tau_\ell}

\def\lamv{\lambda_v}
\def\lams{\lambda_\ms}
\def\lame{\lambda_\me}
\def\lami{\lambda_\mi}
\def\laml{\lambda_\ell}

\def\as{\alpha_\ms}
\def\ae{\alpha_\me}
\def\ai{\alpha_\mi}
\def\al{\alpha_\ell}

\def\H{{\cal H}}

\def\Ee{ {{\cal E}}_\me}
\def\Ei{ {{\cal E}}_\mi}
\def\Es{ {{\cal E}}_\ms}

\def\Xi{ X_\mathrm{i}}

\def\muc{\mathrm{uc}}

\def\dotV{{\dot{V}}}
\def\dotv{{\dot{v}}}

\def\Sr{\ov{r}}
\def\Mr{\wh{r}}

\def\SEs{ {\ov{{\cal E}}}_\ms}
\def\SEe{ {\ov{{\cal E}}}_\me}
\def\SEi{ {\ov{{\cal E}}}_\mi}

\def\Sas{\bar{\alpha}_\ms}
\def\Sae{\bar{\alpha}_\me}
\def\Sai{\bar{\alpha}_\mi}
\def\Mas{\hat{\alpha}_\ms}
\def\Mae{\hat{\alpha}_\me}
\def\Mai{\hat{\alpha}_\mi}

\def\SH{\ov{\H}}
\def\SV{\Ex{\ov{V}}}
\def\SD{\Ex{\ov{\dot{V}}}}

\def\MHe{\Ex{\wh{H_\me}}}
\def\MHi{\Ex{\wh{H_\mi}}}
\def\MH{\wh{\H}}
\def\MV{\Ex{\wh{V}}}

\def\MD{\Ex{\wh{\dot{V}}}}

\def\Thhsx{\Ex{h^2_\ms}_x}

\def\Shhsx{\Ex{\ov{h^2_\ms}}_x}

\def\Mhhsx{\Ex{\wh{h^2_\ms}}_x}
\def\Mhhex{\Ex{\wh{h^2_\me}}_x}
\def\Mhhix{\Ex{\wh{h^2_\mi}}_x}

\def\Tvhex{\Ex{vh_\me}_x}
\def\Tvhix{\Ex{vh_\mi}_x}
\def\Tvhsx{\Ex{vh_\ms}_x}
\def\Svhsx{\Ex{\ov{vh_\ms}}_x}
\def\Svhex{\Ex{\ov{vh_\me}}_x}
\def\Svhix{\Ex{\ov{vh_\mi}}_x}
\def\Mvhsx{\Ex{\wh{vh_\ms}}_x}
\def\Mvhex{\Ex{\wh{vh_\me}}_x}
\def\Mvhix{\Ex{\wh{vh_\mi}}_x}

\def\Tvvx{\Ex{v^2}_x}
\def\Svvx{\Ex{\ov{v^2}}_x}
\def\Svv{\Ex{\ov{v^2}}}
\def\Svvsx{\Ex{\ov{v^2}}_x^\ms}
\def\Svvex{\Ex{\ov{v^2}}_x^\me}
\def\Svvix{\Ex{\ov{v^2}}_x^\mi}

\def\Mvvsx{\Ex{\wh{v^2}}_x^\ms}

\def\Mvv{\Ex{\wh{v^2}}}

\def\Mvve{\Ex{\wh{v^2}}^\me}
\def\Mvvi{\Ex{\wh{v^2}}^\mi}

\def\Tvdx{\Ex{v\dotv}_x}
\def\Mvdx{\Ex{\wh{v\dotv}}_x}
\def\Mvd{\Ex{\wh{v\dotv}}}

\def\Tdhsx{\Ex{\dotv h_\ms}_x}
\def\Tdhex{\Ex{\dotv h_\me}_x}
\def\Tdhix{\Ex{\dotv h_\mi}_x}

\def\Mdhex{\Ex{\wh{\dotv h_\me}}_x}
\def\Mdhix{\Ex{\wh{\dotv h_\mi}}_x}

\def\Sdd{\Ex{\ov{\dot{v}^2}}}
\def\Sddx{\Ex{\ov{\dot{v}^2}}_x}
\def\Sddsx{\Ex{\ov{\dot{v}^2}}_x^s}

\def\Tddx{\Ex{\dot{v}^2}_x}
\def\Mdd{\Ex{\wh{\dot{v}^2}}}
\def\Mddx{\Ex{\wh{\dot{v}^2}}_x}
\def\Mddsx{\Ex{\wh{\dot{v}^2}}_x^s}

\def\Mdde{\Ex{\wh{\dot{v}^2}}^\me}
\def\Mddi{\Ex{\wh{\dot{v}^2}}^\mi}

\begin{document}
\title{Upcrossing-rate dynamics for a minimal neuron model\\ receiving
spatially distributed synaptic drive}
\author{Robert P. Gowers$^{1,2,3}$ and Magnus
  J. E. Richardson$^{1}$\footnote{Correspondence:
    magnus.richardson@warwick.ac.uk}}
\affiliation{
$^1$Warwick Mathematics Institute, University of Warwick, CV4 7AL, United Kingdom, $^2$Institute for Theoretical Biology,
Humboldt-Universit\"at zu Berlin,  10115 Berlin,  Germany, $^3$Bernstein
Center for Computational Neuroscience, 10115 Berlin, Germany
}


\begin{abstract} 
The spatiotemporal stochastic dynamics of the voltage as well as the upcrossing rate are
derived for a model neuron comprising a long dendrite with uniformly
distributed filtered excitatory
and inhibitory synaptic drive. A cascade of ordinary and partial differential equations is
obtained describing the evolution of first-order means and
second-order spatial covariances of the voltage and its rate of change. These quantities provide an
analytical form for the general, steady-state and linear response of the upcrossing rate to
dynamic synaptic input. It is demonstrated that this minimal dendritic model has an
unexpectedly sustained high-frequency response despite synaptic,
membrane and spatial filtering.\\[2mm] \textbf{Reference: Physical
  Review Research (2023)} 
\end{abstract}

\maketitle
\section{Introduction}
Neurons are spatially extended cells receiving a high density of
synapses on their dendrites \cite{Magee2000} and can be
modelled as threshold devices that integrate filtered stochastic input
from presynaptic populations. Over the
last decades there have been significant advances in the
mathematical analysis of neuronal input-output functions,
typically in an approximation in which the cell is treated as isopotential \cite{Brunel2014}. Simultaneously,
there has been growing interest in how spatially induced voltage
differences throughout the dendritic arbour might
support computational capacities beyond the isopotential
approximation. These latter studies have been overwhelmingly simulational \cite{Poirazi2020} due to the
difficulty in accounting for spatial structure and non-linear
filtering. 

There is a relative sparsity of results for stochastic synaptic integration in neurons with explicit
spatial structure
\cite{Tuckwell1983,Manwani1999b,Tuckwell2006,Tuckwell2007,Aspart2016,Gowers2020}. However,
earlier studies of isopotential neurons demonstrate that analytical
statements derived from reduced models provide a general and enduring
framework that are an important guide for biophysically detailed but particular simulational
studies. With this in mind, here a minimal model of spatiotemporal
integration is considered and solved for both the stochastic voltage
and firing-rate dynamics. 

We first derive a set of partial differential equations that describe
the spatiotemporal voltage fluctuations under dendritic integration of
stochastic synaptic drive. We then adapt Rice's level-crossing approximation
\cite{Rice1945}, widely used for isopotential models \cite{Jung1994,Verechtchaguina2006,
  Burak2009,Tchumatchenko2010, Badel2011,Leon2018,Schwalger2021,Sanzeni2022},
to demonstrate that the high-frequency response of the upcrossing rate
exhibits a much weaker effect of the cascade of synaptic, membrane and spatial filtering than might naively be expected.

\section{Model}
The voltage $V(x,t)$ of an infinite dendrite, with a threshold crossing
$V_\mth$ tested at $x\!=\!0$ only, obeys
\be
\hspace{-6mm}\p_t{V}\!&\!=\!&\!\alpha_\ml(E_\ell\!-\!V)\!+\!H_\me(E_\me\!-\!V)\!+\!H_\mi(E_\mi\!-\!V)\!+\!D\p_x^2V  \label{DendVt}
\ee
where the leak and synaptic conductances per unit area have been divided by capacitance per unit area to give rate-like quantities $\alpha_\ell$,
$H_\ms(x,t)$ and where $E_\ml$, $E_\ms$ are the associated reversal
potentials. We will use the notation $\ms\!=\!\me,~\mi$ throughout to denote excitation or
inhibition, respectively. The diffusive term of constant strength
$D\!=\!\laml^2\alpha_\ml$, where $\laml$ is the electrotonic length,
captures the effect of axial-current flow through the dendritic
core. Structually, the model can be interpreted as a neuron with two long dendrites
stemming from a small soma that has no additional conductance load.

\begin{figure}[t]
\centerline{\includegraphics[scale=0.625]{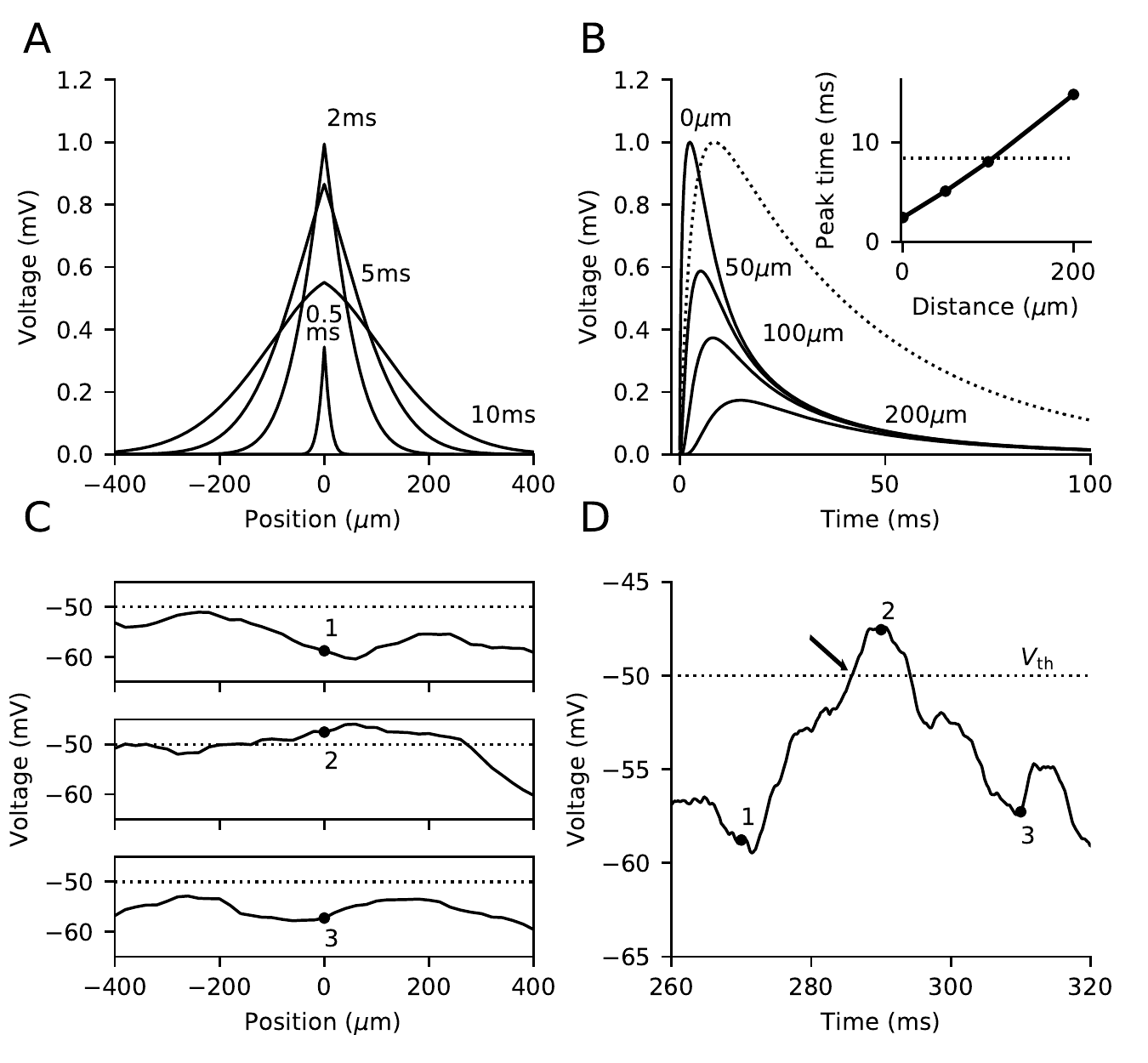}}
\caption{Spatiotemporal voltage profiles for a single synaptic pulse (A-B) and
  widespread stochastic synaptic drive (C-D). (A) Spatial profiles
  for a synaptic pulse at the origin at times marked. (B) Temporal
  profiles at distances marked (corresponding isopotential neuron
  form, dotted line). Inset shows time-to-peak is shorter than the isopotential case (dotted line) for nearby inputs. (C) Spatial profiles of three snapshots separated by
  $20$ms during widespread stochastic synaptic input (threshold
  $V_\mathrm{th}\!=\!-50$mV, dotted line). (D) Temporal voltage
  profile at $x\!=\!0$. Labelled symbols corresond to those in panel 1C. An upcrossing
  event passing $V_{\mth}$ from below is marked (arrow). Parameters used were $(E_\ell,E_\me,E_\mi)\!=\!(-60,-80,0)$mV;
$(\tau_\me,\tau_\mi)\!=\!(3,10)$ms,
$(\al,\Sae,\Sai)\!=\!(25,5.6,11)$Hz and
$(\laml,\lame,\lami)\!=\!(224,19,64)\mu$m. All simulations were written
in Julia \cite{Bezanson2017} with details provided in Appendix D. Code is provided in the repository \textit{Gowers-Richardson-PRR-2023} at
https://github.com/mje-richardson.}
\label{Fig1}
\end{figure}

The
response to an isolated excitatory synaptic input
$\tau_\me\dot{H_\me}+H_\me\propto\delta(x)\delta(t)$, where $\taue$ is
the excitatory synaptic time constant, is plotted in
Fig. \ref{Fig1}A and \ref{Fig1}B. In the latter panel the
temporal profiles at different distances are compared to that
of an isopotential model where $D\!=\!0$ and $\tau_\me\dot{H_\me}+H_\me\propto\delta(t)$. The time to peak for nearby input is shorter than for the isopotential model and so the
cross-over behaviour (see Fig. \ref{Fig1}B inset) suggests that the
minimal dendritic model might have a more rapid response to synaptic
drive than the isopotential model, despite the additional spatial filtering.

To examine whether this is or is not the case, we developed a model of spatially distributed synaptic
drive with the arrival of presynaptic
spikes approximated as space-time Gaussian white-noise processes $\eta_\ms(x,t)$ filtered at physiological timescales $\tau_\me\!=\!3$ms and $\tau_\mi\!=\!10$ms. Therefore 
\be
\taus\dot{H}_\ms\!&\!=\!&\!\as-H_\ms+\sqrt{\as\lams}\,\eta_\ms(x,t)  \label{DendHt}
\ee
where $\as(t)$ is proportional to the
presynaptic rate and $\lams$ a length constant. The zero-mean white noise has autocovariance
$\Ex{\eta_\ms(x_1,t_1)\eta_\ms(x_2,t_2)}\!=\!\delta(x_1\!-\!x_2)\delta(t_1\!-\!t_2)$. Excitation
and inhibition are considered statistically uncorrelated, though this
can be accomodated within the calculational framework to be presented. The
model (Eqs. \ref{DendVt},\ref{DendHt}) is closely
related to Tuckwell's \cite{Tuckwell2006} but includes multiple synaptic timescales
and dynamic conductances.

The voltage and synaptic state-variables are now resolved into
deterministic (mean) and fluctuating (zero mean) components, for example
$V(x,t)=\Ex{V}(t)+v(x,t)$ where the deterministic parts are temporally
dependent but spatially independent and obey
\be
\hspace*{-5mm}&&\p_t\Ex{V}\!=\!\alpha_\ell\l(E_\ell\!-\!\Ex{V}\r)+\Ex{H_\me}\l(E_\me\!-\!\Ex{V}\r)+\Ex{H_\mi}\l(E_\mi\!-\!\Ex{V}\r) \nonumber
\\
\hspace*{-5mm}&&\taue\p_t\Ex{H_\me}\!=\!\ae\!-\!\Ex{H_\me}  ~\mbox{
  and }~\taui\p_t\Ex{H_\mi}\!=\!\ai\!-\!\Ex{H_\mi}.\label{Denddet} 
\ee
The fluctuating components $v,~h_\me,~h_\mi$ are functions of space and time
and obey the partial-differential equations 
\be 
\hspace*{-8mm}&&\p_t{v}=h_\me \Ee +
h_\mi \Ei-\H v +D\p_x^2v \label{Dendfluc} \\
\hspace*{-8mm}&&\taue\p_t{h}_\me=\sqrt{\ae\lame}\,\eta_\me -
h_\me ~\mbox{ and } ~\taui\p_t{h}_\mi=\sqrt{\ai\lami}\,\eta_\mi - h_\mi  \nonumber
\ee
where $\Es(t)\!=\!(E_\ms\!-\!\Ex{V})$ and
$\H(t)\!=\!\alpha_\ell\!+\!\Ex{H_\me}\!+\!\Ex{H_\mi}$ are spatially independent,
though generally time dependent. Note that in
deriving Eqs. (\ref{Denddet}-\ref{Dendfluc}) we have
dropped relatively less significant terms like $\Ex{vh_\me}$  \cite{Manwani1999b,Richardson2005}  so the voltage has
Gaussian statistics. Fig. \ref{Fig1}C and \ref{Fig1}D provide examples of the spatiotemporal
dynamics and an upcrossing event.

The upcrossing rate \cite{Rice1945} is a non-linear function of two first-order and
three second-order voltage moments
$r_\muc(\Ex{V},\Ex{\dot{V}},\Ex{v^2},\Ex{v\dot{v}},\Ex{\dot{v}^2})$
with the full form provided in Appendix A. The
first-order moments are given by Eq. set (\ref{Denddet}). To obtain the second-order moments we derive partial
differential equations for the same-time space-separated
covariances. Introducing the shorthand
$\Thhsx \!=\!\Ex{h_\ms(x_1,t)h_\ms(x_2,t)}$ where $x\!=\!x_2\!-\!x_1$
we first formally solve for the same-time synaptic autocovariance
\be
\hspace*{-3mm}\Thhsx=\delta(x)\frac{\lams}{\taus^2}\!\int_{-\infty}^{t}\!
dt'e^{-2(t-t')/\taus}\as(t'). \label{heautocov}
\ee
This integral is also the solution of a linear
partial-differential equation for $\Thhsx$ (see Eq. \ref{Dendhshsx}). We can also derive partial-differential equations
for other covariances by taking various moments of equations set
(\ref{Dendfluc}) to give 
\be
\hspace*{-8mm}&&\frac{\taus}{2}\p_ t\Thhsx=
\delta(x)\frac{\as\lams}{2\taus}
-\Thhsx \label{Dendhshsx} \\
\hspace*{-8mm}&&\p_ t\Tvhsx\!=\!\Es\Thhsx\!-\!\l(\H\!+\!\frac{1}{\taus}\r)\!\Tvhsx\!+\!D\p^2_x\Tvhsx \label{Dendvhsx} \\
\hspace*{-8mm}&&\frac{1}{2}\p_t\Tvvx\!=\! \Ee\Tvhex\!\!+\!\Ei\Tvhix\!\!-\!\H\Tvvx\!\!+\!D\p^2_x\Tvvx \label{Dendvvx}
\ee
where additionally we have $\Tvdx\!=\!\p_t\Tvvx/2$. For the autocovariance of $\dotv$
we will need the relation
\be
\Tdhsx=\partial_t\Tvhsx+\Tvhsx/\taus
\ee
derived by multiplying the synaptic conductance equation (\ref{Dendfluc}) by $v$ and taking moments while noting that $\Ex{v\eta_{\ms}}\!=\!0$ due to
  causality. The above relation is used for the autocovariance of the rate-of-change of voltage
\be
\hspace*{-6mm}\Tddx\!&\!=\!&\!\Ee\Tdhex+\Ei\Tdhix-\H\Tvdx\!+\!D\p^2_x
\Tvdx.  \label{Denddvdvx} 
\ee
The covariance equations (\ref{Dendhshsx}-\ref{Denddvdvx}),
with $\ms\!=\!\me,\mi$ provide a feedforward cascade allowing all moment-like
quantities to be derived for the upcrossing dynamics by solving for the $x$ and $t$ dependence and then
setting $x=0$. 

\begin{figure}[t]
\centerline{\includegraphics[scale=0.625]{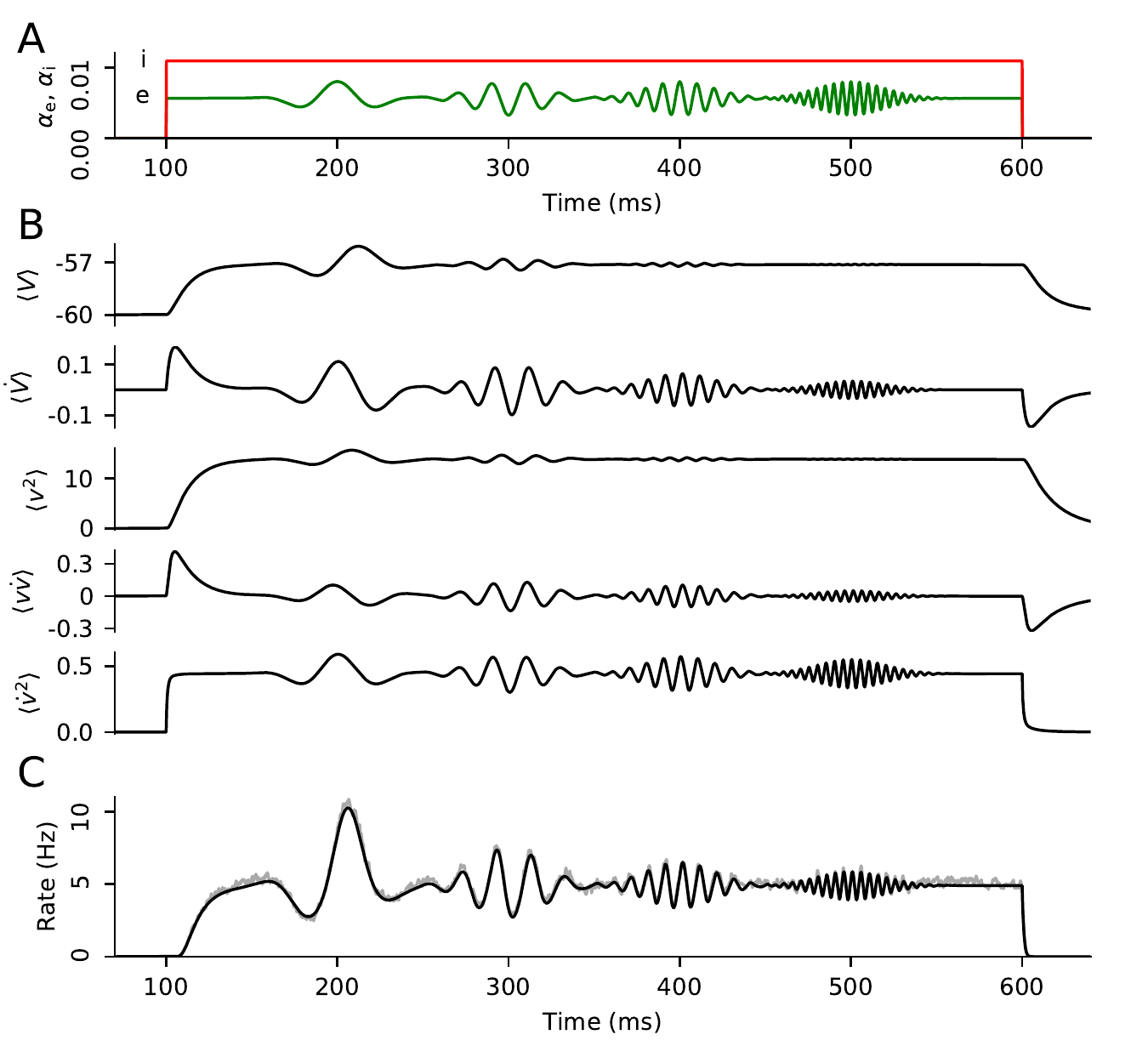}}
\caption{Response to patterned synaptic input (A) comprising step-rate increases in excitatory (green) and inhibitory (red)  drive (same parameters as Fig. 1C) with excitatory
  chirps at $20$, $50$, $100$, $200$Hz. (B) First and second-order voltage moments with those containing a voltage derivative showing stronger responses
  at higher frequencies. (C). The upcrossing rate is a non-linear
  function of the various moments (see Appendix A) and also shows a relatively sustained response at higher
  frequencies, despite the filtering from synapses, spatial spreading and the
  membrane time constant. The mathematical form of the patterned input
is provided in Appendix D.}
\label{Fig2}
\end{figure}

It should be noted that these equations are valid for arbitrary
presynaptic rate dynamics and are not linear approximations. An
example of the response to changes in the presynaptic rates comprising onset/offset and multiple
frequency components is provided in Fig. \ref{Fig2}. It can be seen
that moments including $\dot{V}$ or $\dot{v}$ have sustained
responses at higher frequencies. 

\section{Steady-state properties}
Before calculating frequency-dependent properties, we first
derive forms for the different spatial covariances and moments
required for the steady-state upcrossing rate. The notation
$\bar{Q}$ is used for the steady-state value of a quantity $Q(t)$. 

\begin{figure*}[t]
\centerline{
\includegraphics[scale=0.625]{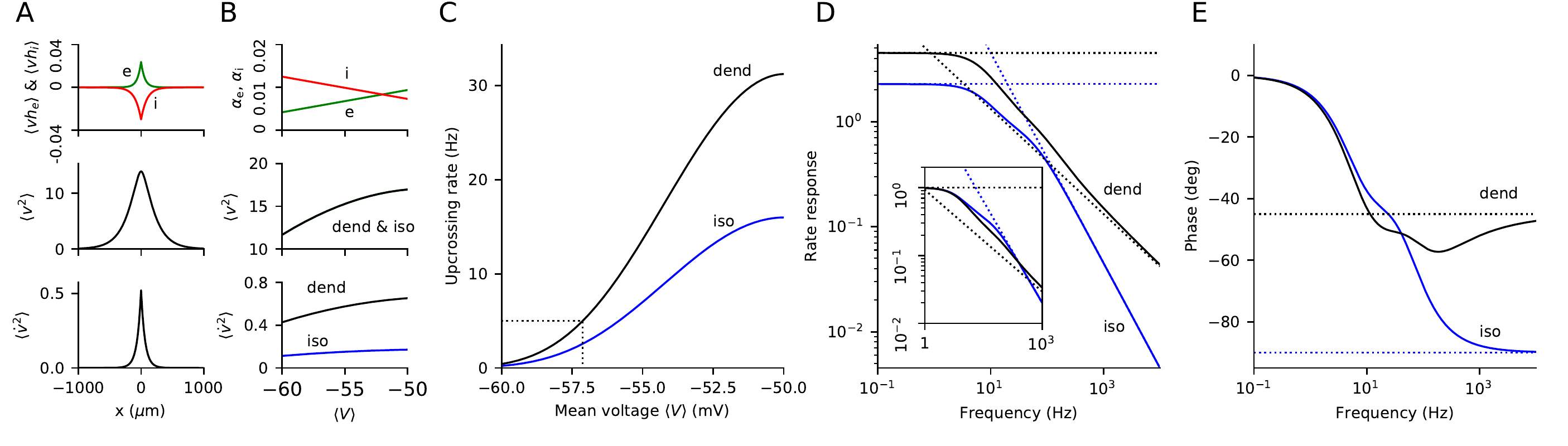}
}
\caption{Steady-state (A-C) and upcrossing-rate response (D-E) showing
  a weakly attenuated response at high
  frequencies. (A) Steady-state spatial covariances of
  synaptic and voltage variables. (B) Steady-state synaptic drive
  covaried to provide a particular mean voltage (x-axis) at fixed
  conductance levels. For an isopotential neuron with matched
  voltage mean, variance and conductance a difference in the rate of change of voltage is seen (lower
  panel, blue). (C) Steady-state upcrossing rate as a function of mean
  voltage for the dendritic (black) and isopotential model (blue). (D)
  Upcrossing-rate response by frequency normalised by $\Mae$. Note that the dendritic-model
  response shows qualitatively weaker attenuation at high-frequency $\sim\!1/\sqrt{i\w}$
  than the reference isopotential model $\sim\!1/i\w$. Inset shows same curves normalised at zero
  frequency in which it is seen that the response of the
  dendritic and isopotential models are broadly similar even over
  moderate frequencies despite the
  additional spatial filtering. (E) Upcrossing phase as a function of
  frequency with a $-45^\circ$ asymptote for the dendritic case and
  $-90^\circ$ for the isopotential model. Parameters used are
  the same as Fig. 1.}
\label{Fig3}
\end{figure*}

The steady-state means are calculated using $\Ex{\bar{H}_{\ms }}\!=\!\Sas$ for
the two synaptic conductances. These give the steady-state average voltage as the
standard weighted average of reversal potentials $\SV\!=\!(\al E_\ell+\Sae E_\me+\Sai E_\mi)\tauv$
where $1/\tauv\!=\!\SH\!=\!\alpha_\ell\!+\!\Sae\!+\!\Sai$. For the steady-state fluctuating components, it proves convenient to introduce an effective space
constant $\lamv$ defined through $\lamv^2\!=\!D\tauv$. We note that the steady-state synaptic conductance fluctuations in
Eq. (\ref{Dendhshsx}) are delta-correlated in space
$\Shhsx\!=\!\delta(x)\Sas/2\taus$ and so
when substituted into the steady-state version of Eq. (\ref{Dendvhsx})
will provide a gradient condition on $\Svhsx$ at
$x\!=\!0$. Given $\psi=\psi_0 e^{-|x|k}$ solves $\psi''=k^2\psi-2k\delta(x)\psi_0$  we have
\be
\Svhsx\!=\!\frac{\SEs}{4\taus}\Sas\tauv\frac{\lams}{\lamv}\sqrt{\frac{\taus}{\tauv+\taus}}e^{-|x|k_s} \label{Dendvhscovar0}
\ee
where $k_s^2\lamv^2\!=\!(\tauv+\taus)/\taus$. An illustration for
excitation and inhibition is
provided in Fig. \ref{Fig3}A,
upper panel. The equation for the steady-state voltage
autocovariance is separated into excitatory and inhibitory components
$\Svvx\!=\!\Svvex+\Svvix$ and solved similarly (see Appendix B) 
\be
\hspace*{-2mm}\Svvsx=\frac{\SEs^2}{4}\Sas\tauv\frac{\lams}{\lamv}\l(\!e^{-|x|k_v}\!-\!\sqrt{\frac{\taus}{\tauv+\taus}}e^{-|x|k_\ms}\!\r) \label{Dendvvcovar0}
\ee
where $k_v\!=\!1/\lamv$. Unlike the
covariance between voltage and a synaptic drive, the voltage
autocovariance has zero gradient at the origin (see Fig. \ref{Fig3}A,
middle panel). The final quantity needed for the steady-state upcrossing
rate is the autocovariance of $\dotv$ that takes the form
$\Sddx=\SEe\Svhex/\tau_\me+\SEi\Svhix/\tau_\mi$.
Each synaptic component of this quantity is easily expressed using the
second of the two results in Eq. (\ref{Dendvhscovar0}) and so
\be
\Sddsx&=&\frac{\SEs^2}{4\tau_\ms^2}\Sas\tauv\frac{\lams}{\lamv}\sqrt{\frac{\taus}{\taus+\tauv}}e^{-|x|k_\ms} \label{Denddvdvcovar0}
\ee
with an illustration provided in Fig. \ref{Fig3}A, lower panel. The result for $\SV\!$ and
Eqs (\ref{Dendvvcovar0}-\ref{Denddvdvcovar0}) evaluated at $x\!=\!0$
provide the quantities needed for the steady-state
upcrossing rate (see Fig. \ref{Fig3}C).

\section{Firing-rate response}
We now derive the frequency-dependent response by considering weak sinusoidal modulations of the incoming
excitatory synaptic rate $\ae(t)\!=\!\Sae\!+\!\Mae e^{i\w t}$ and
expand all state variables to leading order in $\Mae$. We will use the notation for some quantity
$Q(t)\!=\!\bar{Q}+\hat{Q}e^{i\w t}$ with $\bar{Q}$ the steady-state value
and $\hat{Q}$ the linear response proportional to $\Mae$. At this
level, the upcrossing rate response $\hat{r}_\mathrm{uc}$ will be a linear function
of the modulated moments (see Appendix A). 

The strategy is similar to
that taken for the steady state but with
Eqs. (\ref{Dendhshsx}-\ref{Denddvdvx}) solved 
in the frequency domain. The calculation is algebraically
lengthy so here we provide the high-frequency asymptotics with the
full forms given in Appendix B. At the mean level
\be
\MV\sim\frac{\SEe\Mae}{(i\w)^2\taue} &\mbox{and}& \MD\sim\frac{\SEe\Mae}{i\w\taue} 
\ee
so the rate-of-change of the average voltage is the dominant
deterministic contribution to the upcrossing-rate
response at higher frequencies. 

For the fluctuating components, the driving excitatory synaptic
modulation is again delta-correlated in space
$\Mhhex=\delta(x)\Mae\lame/2\taue(1+i\w\taue/2)$
but with a frequency-dependent amplitude due to synaptic
filtering. Using this result, solving for the response of the voltage and synaptic
covariances, the high-frequency
asymptote of the voltage variance is found:
\be
\hspace*{-5mm}\Mvv\sim\!\!&\!\!&\!\!-\frac{2\Mae\tauv}{i\w\taue}\frac{\Svv}{i\w\tauv}
\ee
and so decays as $1/\w^2$. From
$\Mvd\!=\!(i\w/2)\Mvv$ this also gives the weaker decay of $\Mvd\sim 1/i\w$. Finally, the
asymptote of the variance of the rate-of-change of voltage 
\be
\Mdd\sim\Mae\tauv\frac{\SEe^2}{2\taue^2}\frac{\lame}{\lamv}\frac{1}{\sqrt{2i\w\tauv}}
\ee
can be seen to have the weakest decay and therefore is dominant at
high frequencies. 

This is the key and somewhat surprising result for the dynamics of the
dendritic model: the high-frequency asymptotics decay as
$1/\sqrt{i\w}$ and, through its linear dependence on $\Mdd$ as seen in
Eq. (\ref{Mruc}) of Appendix A, so also must the high-frequency response of
the firing rate in the upcrossing approximation
\be
\frac{\Mr_\muc}{\Sr_\muc}\sim \Mae\tauv\frac{\SEe^2}{4\taue^2\Sdd}\frac{\lame}{\lamv}\frac{1}{\sqrt{2i\w\tauv}}.
\ee
This can be contrasted to the result for the isopotential point-neuron
model that has an upcrossing response decaying as $1/i\w$ at higher
frequencies (see reference \cite{Badel2011}
and Appendix C). In Figs. \ref{Fig3}D and \ref{Fig3}E an illustration of the
amplitude and phase of the response is shown. These frequency-domain results are
compatible with the earlier observation in Fig. \ref{Fig1}B that EPSPs
on a dendrite can be sharper in time than for an isopotential model.

\section{Discussion}
The analyses presented here are predicated on a number of biophysical
approximations and therefore should be considered as providing the basis
for future refinement. 

Firstly, the membrane model does not include voltage-gated currents such as
the h-current that can affect low frequency components of the firing-rate response. These
could be included using a quasi-active membrane approximation
\cite{Koch1984, Coombes2007} with additional state variables coupled
to the voltage dynamics. 

The minimal model presented
here also approximates spatial extent as infinite (valid for dendrites
significantly longer than the effective electrotonic length
$\lambda_v$), is homogeneous and has no increased conductance at the
position $x\!=\!0$ of the nominal soma. Recent analysis
\cite{Gowers2020} showed significant effects of geometry on the
functional forms of steady-state upcrossing rates. The derivation
of Eqs. (\ref{Dendhshsx}-\ref{Denddvdvx}) rely on a long, homogeneous approximation and so adaptation of the method to more realistic
geometries might be a technical challenge, though the spatial-mode
expansion  technique used by Tuckwell \cite{Tuckwell2006} is a
potential strategy to account for closed-end effects.

A number of
approximations of the synaptic drive have been made including the Gaussian approximation of
finite-amplitude shot noise. This typically has validity when
statistically independent, high-rate, low-ampltude inputs are
summed.  Given the distinct response seen in isopotential neurons when shot noise is included \cite{Droste2017,Richardson2018}, a
worthwhile extension would be to examine
finite-amplitude effects on the dynamics. This is particularly
important for spatiotemporal integration as the relative number of
summed inputs within an effective electrotonic length will be less than the global input into an isopotential model.

Finally, though widely used in neuroscience, the upcrossing approximation should be
critically evaluated in this spatial context and compared to biophysical models
of spike generation. Rapid responses have already been
identified in these models due to spiking non-linearities or somatic-dendritic
coupling \cite{Fourcaud2005,Naundorf2006,
  Ilin2013,Eyal2014,Ostojic2015,Doose2017}. Extensions of the current
study could examine the high-frequency response when both stochastic
spatiotemporal integration and non-linearities known to affect the
rapidity of action-potential
generation are combined.

\begin{acknowledgments}
\ni We would like to thank thank Nicolas Brunel and Benjamin Lindner for their useful comments on an earlier version of this
manuscript. We also acknowledge funding from the Engineering and Physical Sciences
Research Council funding under Grant No. EP/L015374/1 to RPG.
\end{acknowledgments}

\section*{APPENDIX A. Upcrossing-rate dynamics}
\vspace*{-3mm}
The time-dependent rate $r_\muc(t)$ that a fluctuating membrane
voltage $V$ crosses a threshold $V_\mth$ from below is
considered. Following Rice \cite{Rice1945}, this can be written as
\be
r_\muc(t)=\int_0^\infty d\dotV\dotV \Psi(V_\mth,\dotV)
\ee
where $\dot{V}$ is the rate-of-change of voltage and $\Psi(V,\dot{V})$ is the
joint probability density. The derivations that will be used for the dynamics,
steady state and linear response were given by Badel
\cite{Badel2011} in the context of a related isopotential neuronal model. We
repeat that derivation and provide intermediate steps for transparency.

It is first convenient to expand the voltage and its rate of change around their time-dependent mean values
$\Ex{V}$ and $\Ex{\dot{V}}$ so the fluctuating excesses $v$ and
$\dot{v}$ have zero mean: for example,
$V(t)=\Ex{V(t)}+v$. Writing the joint distribution for $v$ and
$\dot{v}$ as the conditional distribution $\psi(\dotv|v)$  multiplied by the marginal voltage density $\phi(v)$ we have
\be
r_\muc(t)=\phi(v_\mth)\int_{-\Ex{\dotV}}^\infty
d\dotv(\Ex{\dotV}+\dot{v} )\psi(\dotv|v_\mth) 
\ee
where $v_\mth(t)=V_\mth-\Ex{V}$. For the Gaussian-distributed voltages considered in this paper, the
distributions can be written 
\be
\phi(v)&=&\frac{1}{\sqrt{2\pi\Ex{v^2}}}\exp\l(-\frac{v^2}{2\Ex{v^2}}\r)
~~\mbox{ and } \\
 \psi(\dot{v}|v)&=&\frac{1}{\sqrt{2\pi
  s^2}}\exp\l(-\frac{(\dotv-\kappa v)^2}{2s^2}\r)
\ee
where the variances $\Ex{v^2}$, $\Ex{\dot{v}^2}$, covariance
$\Ex{v\dot{v}}$ and other parameters $\kappa=\Ex{v\dotv}/\Ex{v^2}$ and
$s^2=\Ex{\dotv^2}-\kappa^2\Ex{v^2}$ are all potentially time dependent. Using these results we get for the upcrossing rate 
\be
r_\muc(t)=\frac{1}{2\pi}\sqrt{\frac{s^2}{\Ex{v^2}}}\,e^{-v_\mth^2/2\Ex{v^2}}\int_{-\beta}^\infty
du (2u+2\beta)e^{-u^2} 
\ee
where $\beta=(\Ex{\dotV}+\kappa v_\mth)/\sqrt{2s^2}$. The integral can be rewritten in terms of Gaussians and the error function 
\be
\hspace*{-2mm}r_\muc(t)\!=\!\frac{1}{2\pi}\sqrt{\frac{s^2}{\Ex{v^2}}}e^{-v_\mth^2/2\Ex{v^2}}\!\l(\!
e^{-\beta^2}\!\!+\!\sqrt{\pi}\beta[1\!+\!\erf(\beta)]\!\r) 
\ee
which is identical to the result arrived at by Badel
\cite{Badel2011}. An example of the upcrossing rate in a regime that
is non-linear in the synaptic driving terms is illustrated in Fig. \ref{Fig2}C (lower panel).

\subsection*{Steady-state upcrossing rate}
\vspace*{-3mm}
For a quantity $Q(t)$ evaluated in the steady state we use the
notation $\ov{Q}$. The steady-state upcrossing rate simplifies because $\SD\!=\!0$ and
$\Svd=\partial_t\Svv/2\!=\!0$ so that $\ov{\beta}\!=\!0$ and
$\ov{s^2}\!=\!\Sdd$ giving
\be
\bar{r}_\muc=\frac{1}{2\pi}\sqrt{\frac{\Ex{\ov{\dotv^2}}}{\Ex{\ov{v^2}}}}\exp\l(-\frac{\bar{v}_\mth^2}{2\Svv}\r)
\ee
where $\bar{v}_\mth\!=\!V_\mth\!-\!\SV$. Figure \ref{Fig3}C provides
an illustration of the steady-state upcrossing rate.

\subsection*{Linear response of the upcrossing rate}
\vspace*{-3mm}
We now consider a weak harmonic modulation of the incoming presynaptic rates. This will induce weak
modulations, with some amplitude and phase shift, in any dependent
quantity $Q(t)$ that we can conveniently write in complex form
$Q(t)=\ov{Q}+\wh{Q}e^{i \w t}$.  Before expanding the upcrossing form,
let us examine some
of the component quantities. For $\beta$ and $s^2$ we have
\be
\hspace*{-5mm}\wh{\beta}=\frac{1}{\sqrt{2\Sdd}}\l(\MD+\ov{v}_\mth\frac{\Mvd}{\Svv}\r)
&\mbox{
and }&\wh{s^2}=\Mdd.
\ee
Then, for the upcrossing rate itself, we get that
the ratio of the modulation to the steady-state rate is \cite{Badel2011}
\be
\hspace*{-10mm}&&\frac{\hat{r}_\muc}{\bar{r}_\muc}=\frac{\bar{v}_\mth}{\Svv}\MV+
\sqrt{\frac{\pi}{2\Sdd}}\MD \nonumber\\
\hspace*{-10mm}&&+\frac{1}{2}\frac{\Mvv}{\Svv}\l(\frac{\bar{v}_\mth^2}{\Svv}-1
\r)+\sqrt{\frac{\pi}{2\Svv}}\bar{v}_\mth\frac{\Mvd}{\Svv}+\frac{1}{2}\frac{\Mdd}{\Sdd}. \label{Mruc}
\ee
In the above equation, the first two terms provide the deterministic
contribution and the last three terms are contributions from modulated
fluctuating quantities. The amplitude and phase of the upcrossing
linear response is shown in Figs. \ref{Fig3}D and \ref{Fig3}E, respectively.

\section*{APPENDIX B. Dendritic model}
\vspace*{-3mm}
The differential equations for the deterministic (mean) components in
Eqs. set (\ref{Denddet}) and the partial differential equations for
the fluctuating components  (covariances) in Eqs.
(\ref{Dendhshsx}-\ref{Denddvdvx}) completely determine the moment
dynamics in the Gaussian approximation of the model. These
equations are driven by the rate-like
terms $\alpha_\me(t)$, $\alpha_\mi(t)$ that are proportional to the presynaptic excitatory and inhibitory rates. Also
appearing in the equations are the total conductance
$\H(t)=\alpha_\ell+\Ex{H_\me}+\Ex{H_\mi}$ and electromotive forcing
terms $\Es(t)=E_\ms-\Ex{V}$. Together, these equations represent a
feedforward cascade that provide all the required quantities needed
for the upcrossing rate. 

There are
a number of approaches that can be taken to find the solution of these equations in the
steady state or at the linear-response level. For example direct
solution in space using substitution for the inhomogeneous components or using spatial
Fourier transforms. Here we will use the former real-space approach
and therefore the following result
will often be useful
\be
\partial_x^2\psi=k^2\psi-2 k\delta(x)\psi_0~\mbox{ with soln. }~
\psi=\psi_0e^{-|x|k}. \label{DxxSolnA}
\ee

\subsection*{Steady state: dendritic model}
\vspace*{-3mm}
We first derive the various same-time space-separated covariances in
the steady state as these will be used to calculate the time
dependence.

\ni\textit{Synaptic autocovariances
  $\Shhsx$}. From Eq. (\ref{Dendhshsx}), these are simply delta-correlated in space
\be
\Shhsx=\delta(x)\frac{\Sas\lambda_\ms}{2\tau_\ms} .
\ee
\ni\textit{Voltage and synaptic covariances
  $\Svhsx$}. In the steady-state, Eq. (\ref{Dendvhsx}) reduces to 
\be
D\frac{\p^2}{\p
  x^2}\Svhsx=\l(\frac{1}{\taus}+\frac{1}{\tauv}\r)\Svhsx-\SEs\Shhsx. \label{DBvhsxA}
\ee
Remembering that $D\!=\!\lamv^2/\tauv$ and looking at the form of
Eq. (\ref{DxxSolnA}) identifies
$\lamv^2k_\ms^2\!=\!(\tauv\!+\!\taus)/\taus$. From the prefactor of
the delta-correlated inhomogeneous term, the solution must therefore be
\be
\Svhsx&=&\frac{1}{2\lambda_v
  k_{\ms}}\left[\frac{\SEs}{2\tau_\ms}\Sas\tauv\frac{\lambda_\ms}{\lamv}\right]e^{-|x|k_{s}}
\nonumber \\
&=&\frac{\SEs}{4\taus}\Sas\tauv\frac{\lams}{\lamv}\sqrt{\frac{\taus}{\taus+\tauv}}e^{-|x|k_\ms}. 
\ee

\ni\textit{Voltage autocovariance $\Svvx$}. There are two
inhomogeneous terms in its equation 
\be
D\frac{\p^2}{\p x^2}\Svvx=\frac{1}{\tauv} \Svvx-\SEe\Svhex-\SEi\Svhix \label{DBvvxA}
\ee
so it can be resolved into
$\Svvx=\Svvex+\Svvix$. Trying $\Svvsx=\psi_\ms+c_\ms\Svhsx$ and
using Eq. (\ref{DBvhsxA}) to remove the double derivative requires
setting $c_\ms=-\taus \SEs$ to cancel the inhomogeneous term. This
leaves
\be
D\frac{\p^2}{\p x^2}\psi_\ms=\frac{1}{\tauv}\psi_\ms-\SEs^2\taus
\Shhsx. 
\ee
Introducing $\kv^2=1/\lamv^2$ the solution for $\psi_\ms$ is
\be
\psi_\ms&=&\frac{\SEs^2}{4}\Sas\tauv\frac{\lame}{\lamv}e^{-|x|\kv}.
\ee
Putting these forms in $\Svvsx=\psi_\ms+c_\ms\Svhsx$ gives
\be
\hspace*{-4mm}\Svvsx\!=\!\left(
  \frac{\SEs^2}{4}\Sas\tauv\frac{\lams}{\lamv}\right)\!\left(e^{-|x|\kv}
-\!\sqrt{\frac{\taus}{\tauv+\taus}}e^{-|x|\ks}\right)\!. \label{SvvsxA}
\ee
It can be noted that this gives the voltage autocovariance a zero gradient at $x\!=\!0$.

\ni\textit{Rate-of-change-of-voltage autocovariance $\Sddx$}. In the
steady-state this is simply
\be
\Sddx&=&\frac{\SEe}{\taue}\Svhex+\frac{\SEi}{\taui}\Svhix 
\ee
where the forms for $\Svhsx$ have already been given above.

\subsection*{Deterministic weak oscillations: dendritic model}
\vspace*{-3mm}
Modulation of the excitatory presynaptic drive $\Mae$ only is
considered, so the modulated inhibitory drive is
zero $\Mai\!=\!0$. With this in mind, expanding the deterministic equations (Eq. set
\ref{Denddet}) at the level of the linear response to excitatory oscillations gives the following quantities of interest 
\be
\MHe=\MH=\frac{\Mae}{1+i\w\taue} ~\mbox{ and
} ~\MHi=0 
\ee
so
\be
\MV=\frac{\SEs\MHe}{1/\tauv+i\w}=\frac{\Mas\tauv\SEs}{(1+i\w\taus)(1+i\w\tauv)}
\ee
and the modulated rate-of-change of the voltage is given by
$\MD=i\w\MV$. Note also that $\MEs=-\MV$ for $\ms=\me,\mi$.

\subsection*{Weak oscillations and fluctuations: dendritic model}
\vspace*{-3mm}
We present the modulated moment derivations in the order of the
cascade of equations, remembering again that for modulated excitatory
drive only we have $\Mai\!=\!0$ throughout.

\ni\textit{Synaptic autocovariances $\Mhhsx$}. These are delta-correlated
\be
\hspace*{-5mm}\Mhhex=\delta(x)\frac{\Mae\lame}{2\taue} \frac{1}{1+i\w\taue/2} &\mbox{
and }& \Mhhix=0.
\ee

\ni\textit{Voltage and synaptic covariances $\Mvhsx$}. This obeys
\be
\hspace*{-5mm}D\partial_x^2 \Mvhsx&=&\left(i\w
  +\frac{1}{\tauv}+\frac{1}{\taus}\right)\Mvhsx \nonumber \\
\hspace*{-5mm}&&+\MH\Svhsx-\SEs\Mhhsx-\MEs\Shhsx \label{DMvhsxA}.
\ee 
We then use a substitution of the form $\Mvhsx\!=\!\psi_\ms\! +\!
a_\ms\Svhsx$ and use the result of Eq. (\ref{DBvhsxA}) to remove the
double spatial derivative on $\Svhsx$. Setting $a_\ms=-\MH/i\w$ then removes the
remaining inhomogeneous term in $\Svhsx$ to leave
\be
\partial_x^2 \psi_\ms=\qs^2\psi_\ms-\frac{\SEs}{D}\Mhhsx-\frac{1}{D}\l(\MEs-a_\ms\SEe\r)\Shhsx
\nonumber 
\ee
where $\lamv^2\qs^2\!=\!(1+\tauv/\taus+i\w\tauv)$. This is
straightforwardly solved and, when combined
with the other inhomogeneous term, gives
\be
\Mvhsx\!&\!=\!&\!
\frac{1}{4\taus}\frac{\lams}{\lamv}\!\l(\frac{\SEs\Mas\tauv}{1+i\w\taus/2}\!+\!\MEs\Sas\tauv\!+\!\frac{\MH\SEs}{i\w}\Sas\tauv\r)\!
\frac{e^{-|x|\qs}}{\qs\lamv} \nonumber \\
&&\!-\frac{\MH}{i\w}\Svhsx.
\ee
Note that we would have $\Mai\!=\!0$ in the first term for the inhibitory form $\Mvhix$. 

\ni\textit{Voltage autocovariance $\Mvvx$}. This obeys
\be
\hspace*{-10mm}&&D\partial_x^2 \Mvvx=\left(\frac{i\w}{2}
  +\frac{1}{\tauv}\right)\Mvvx+\MH\Svvx \nonumber \\
\hspace*{-10mm}&&\hspace*{5mm}-\SEe\Mvhex-\MEe\Svhex-\SEi\Mvhix-\MEi\Svhix.
\ee
We can separate this into components for excitation and inhibition,
each of which satisfies
\be
D\partial_x^2 \Mvvsx&=&\left(\frac{i\w}{2}
  +\frac{1}{\tauv}\right)\Mvvsx+\MH\Svvsx \nonumber \\
&&-\SEs\Mvhsx-\MEs\Svhsx.
\ee
These can be solved by substituting
$\Mvvsx=a\Svvsx+b_\ms\Mvhsx+c_\ms\Svhsx+\psi_\ms$ giving
\be
&&aD\partial_x^2 \Svvsx +b_\ms D\partial_x^2\Mvhsx+c_\ms D\partial_x^2\Svhsx+D\partial_x^2\psi_\ms =\nonumber \\
&&\left(\frac{i\w}{2}
  +\frac{1}{\tauv}\right)\l(a\Svvsx+b_\ms\Mvhsx+c_\ms\Svhsx+\psi_\ms
\r)+ \nonumber \\
&&\MH\Svvsx-\l[\SEs\Mvhsx+\MEs\Svhsx\r].
\ee
We now replace the double spatial derivatives $D\partial_x^2 \Svvsx$, $D\partial_x^2
\Mvhsx$ and $D\partial_x^2 \Svhsx$ using Eq. (\ref{DBvvxA}) resolved into $\ms$-dependent components for $\Svvsx$ as well as Eqs. (\ref{DMvhsxA}, \ref{DBvhsxA})
respectively for $\Mvhsx$ and $\Svhsx$, to give
\be
&&\hspace{-5mm}a\l[\frac{1}{\tauv} \Svvsx-\SEs\Svhsx\r]+\nonumber \\
&&\hspace{-5mm}b_\ms\l[ \left(i\w
  \!+\!\frac{1}{\tauv}\!+\!\frac{1}{\taus}\right)\Mvhsx\!+\!\MH\Svhsx\!-\!\SEs\Mhhsx\!-\!\MEs\Shhsx\r]\nonumber
\\
&&\hspace{-5mm}+c_\ms\l[\left(\frac{1}{\tauv}+\frac{1}{\taus}\right)\Svhsx-\SEs\Shhsx\r]+D\partial_x^2\psi =\nonumber \\
&&\hspace{5mm}\left(\frac{i\w}{2}
  +\frac{1}{\tauv}\right)\l(a\Svvsx+b_\ms\Mvhsx+c_\ms\Svhsx+\psi_\ms
\r)\nonumber \\
&&\hspace{15mm}+\MH\Svvsx-\SEs\Mvhsx-\MEs\Svhsx.
\ee
We then set $a$, $b_\ms$ and $c_\ms$ to remove the
inhomogeneous terms in $\Svvsx$, $\Mvhsx$ and $\Svhsx$
respectively:
\be
&&a=-\frac{2\MH}{i\w},~~b_\ms=\frac{-\SEs\taus}{1+i\w\taus/2},\nonumber
\\
&&\hspace{10mm}~\mbox{ and }~
c_\ms=\frac{\taus(a\SEs-\MEs-b_\ms\MH)}{1-i\w\taus/2} \label{abscs}
\ee
and leave an equation for $\psi_\ms$ of the form
\be
&&D\partial_x^2\psi_\ms =\left(\frac{i\w}{2}
  +\frac{1}{\tauv}\right)\psi_\ms \nonumber \\
&& +b_\ms\SEs\Mhhsx+b_\ms\MEs\Shhsx+c_\ms\SEs\Shhsx.
\ee
This equation has solution
\be
\psi_\ms&=&-\frac{1}{4\taus}\frac{\lams}{\lamv}\l[b_\ms\l(\frac{\SEs\Mas\tauv}{1+i\w\taus/2}+\MEs\Sas\tauv\r)+c_\ms\SEs\Sas\tauv\r]\nonumber
\\
&&\hspace{10mm}\times\frac{e^{-|x|\qv}}{\lamv\qv} \label{phisA}
\ee
which together with the other inhomogeneous forms
in $\Mvvsx=a\Svvsx+b_\ms\Mvhsx+c_\ms\Svhsx+\psi_\ms$ completes the solution for one
synaptic component of the
modulated variance.

\ni\textit{Rate-of-change of voltage autocovariance $\Mddx$}. This has form
\be
\Mddx&=&\SEe\Mdhex+\MEe\Sdhex+\SEi\Mdhix+\MEi\Sdhix\nonumber \\
&&-\frac{1}{\tauv}\Mvdx+D\partial_x^2 \Mvdx.
\ee
We again separate out the solution in terms of the
components involving excitation and inhibition
\be
\Mddsx&=&\frac{\SEs}{\taus}(1+i\w\taus)\Mvhsx+\frac{\MEs}{\taus}\Svhsx
\nonumber \\
&&-\frac{i\w}{2\tauv}\Mvvsx+\frac{i\w }{2}D\partial_x^2 \Mvvsx
\ee
where we have also made use of the simplifying relations for $\Ex{\dot{v}h_\ms}$ and
$\Ex{v\dot{v}}$ in the steady-state and linear-response levels. We now substitute for the following term
\be
D\partial_x^2 \Mvvsx&=&\left(\frac{i\w}{2}
  +\frac{1}{\tauv}\right)\Mvvsx \nonumber \\
&&+\MH\Svvsx-\SEs\Mvhsx-\MEs\Svhsx
\ee
and tidy things up to get 
\be
\Mddsx&=&\frac{\SEs}{\taus}\l(1+\frac{i\w\taus}{2}\r)\Mvhsx+\frac{\MEs}{\taus}\l(1-\frac{i\w\taus}{2}\r)\Svhsx\nonumber
\\ &&+\l(\frac{i\w}{2}\r)^2\Mvvsx+\frac{i\w}{2}\MH\Svvsx
\ee
which is expressed in terms of quantities already derived.

\subsection*{Low-frequency limit: dendritic model}
\vspace*{-3mm}
In the limit $\w\!\to\!0$, the various frequency-dependent quantities $\wh{Q}(\w)$
can be obtained by taking derivatives of corresponding steady-state quantities with respect to $\Sae$
\be
\lim_{\w\to 0} \wh{Q}=\Mae\frac{d}{d\Sae} \ov{Q}
\ee
where it should be remembered that $\tauv, \lamv,
\SEe, \SEi$ all depend on $\Sae$. The following results are useful
\be
\hspace*{-12mm}&&\frac{d}{d\Sae}\frac{1}{\tauv}=1,~~~\frac{d}{d\Sae}\tauv=-\tauv^2,\nonumber
\\
\hspace{-12mm}&&\frac{d}{d\Sae}\SEs=-\frac{d}{d\Sae}\SV=-\tauv\SEe
~\mbox{ and }~ \frac{d}{d\Sae}\frac{1}{\lamv}\!=\!\frac{1}{2}\frac{\tauv}{\lamv}.
\ee
It is also useful to introduce the following definition and its derivatives
\be
\hspace*{-5mm}x_s=\frac{\taus}{\taus+\tauv} &\mbox{ so }&
\frac{d}{d\Sae}x_s=\tauv x_s(1-x_s)=\frac{\tauv^2}{\taus}x_s^2.
\ee
Finally, note that because $\dot{V}$ or $\Ex{v\dot{v}}$ are both
complete temporal derivates, their temporal Fourier transforms vanish
in the $\w\!=\!0$. We now provide the forms of the remaining moments.

\ni\textit{Voltage and synaptic covariance $\Mvhs$}. In terms of
$x_\me$ and $x_\mi$ these can be written
\be
\hspace*{-6mm}\lim_{\w\to 0}\Mvhe&\!=\!&\Mae\tauv\frac{\SEe}{4\taue}\frac{\lame}{\lamv}
\sqrt{x_\me}\l(1-\Sae\tauv\l(1+\frac{x_\me}{2} \r)\r) \nonumber \\
\hspace*{-6mm}\lim_{\w\to 0}\Mvhi&\!=\!&- \Mae\tauv\frac{1}{4\taui}\frac{\lami}{\lamv}
\sqrt{x_\mi}\Sai\tauv\l(\SEe +\SEi\frac{x_\mi}{2} \r).
\ee

\ni\textit{Voltage variance $\Mvv$}. We can split this term into excitatory and
inhibitory components and use the same definitions for $x_\me$ and
$x_\mi$ as above
\be
&&\hspace*{-10mm}\lim_{\w\to 0}\Mvve=
\Mae\tauv\frac{\SEe^2}{4}\frac{\lame}{\lamv}\times\nonumber \\
&&\l( (1-2\Sae\tauv)\l(1-\sqrt{x_\me}\r)-\frac{\Sae\tauv}{2}\l(1-\sqrt{x_\me^3}\r)\r)\nonumber \\
&&\hspace*{-10mm}\lim_{\w\to
  0}\Mvvi=-\Mae\tauv\frac{1}{4}\frac{\lami}{\lamv}\Sai\tauv\times
\nonumber \\
&&\l(2\SEe\SEi\l(1-
\sqrt{x_\mi}\r)   +\frac{\SEi^2}{2}    \l(1-\sqrt{x_\mi^3}\r).
\r)
\ee

\ni\textit{Rate-of-change of voltage variance $\Mdd$}. The excitatory and inhibitory components are proportional to
$\Mvhe$ and $\Mvhi$ so that
\be
\hspace*{-7mm}\lim_{\w\to 0}\Mdde&\!=\!&\Mae\tauv\frac{\SEe^2}{4\taue^2}\frac{\lame}{\lamv}
\sqrt{x_\me}\l(1-\Sae\tauv\l(2+\frac{x_\me}{2} \r)\r) \nonumber \\
\hspace*{-7mm}\lim_{\w\to 0}\Mddi&\!=\!& -\Mae\tauv\frac{1}{4\taui^2}\frac{\lami}{\lamv}
\sqrt{x_\mi}\Sai\tauv\l(2\SEe\SEi +\SEi^2\frac{x_\mi}{2} \r).
\ee

\subsection*{High-frequency asymptotics: dendritic model}
\vspace*{-3mm}
For a modulation of the excitatory component, to leading order, the
deterministic components needed are
\be
\hspace{-15mm}&&\MV\sim\frac{\Mae\SEe}{(i\w)^2\taue},~~\MD\sim
\frac{\Mae\SEe}{i\w\taue}, \nonumber \\
\hspace{-15mm}&&\MHe=\MH\sim\frac{\Mae}{i\w\taue},~~\mbox{ and }~~\MEe=\MEi=-\MV.
\ee
Note that $\MHi=0$ because $\Mai=0$. The dominant contribution to the
deterministic component to the upcrossing rate is therefore $1/i\w$ and
comes from the rate-of-change of voltage term. We now take the covariances in
turn. 

\ni\textit{Voltage and synaptic covariance
  $\Mvhs$}. For the covariances between voltage and synaptic drive we have
\be
\hspace{-15mm}&&\Mvhe\sim\Mae\tauv\frac{\SEe}{2\taue}\frac{\lame}{\lamv}\frac{1}{i\w\taue}\frac{1}{\sqrt{i\w\tauv}}~~\mbox{ and }
\nonumber \\
\hspace{-15mm} &&\Mvhi\sim-\frac{\Mae\tauv}{i\w\taue}\frac{\Svhi}{i\w\tauv}.
\ee

\ni\textit{Voltage variance $\Mvv$ and $\Mvd$}. Examining the
forms of the various terms in Eq. set (\ref{abscs}) we see that $a\sim1/\w^2$, $b_\ms\sim 1/\w$ and
$c_\ms\sim 1/\w^3$. The term multiplying the exponential therefore decays
as $1/\w^{5/2}$ and is less significant that the $a\Svvsx$ term, which
dominates the inhomogeneous parts of the solution. Using the asymptotics
for $a$ then gives 
\be
\hspace{-7mm}\Mvv\sim-\frac{\Mae\tauv}{i\w\tauv}\frac{2}{i\w\taue}\Svv &\mbox{and}&\Mvd\sim-\frac{\Mae}{i\w\taue}\Svv
\ee
where the latter result follows from $\Mvd=(i\w/2)\Mvv$.

\ni\textit{Rate-of-change of voltage variance $\Mdd$}. It is useful to re-arrange the form of this equation so that
\be
\Mddsx&=&\l(\frac{i\w}{2}\r)^2\l[\Mvvsx+\frac{2}{i\w}\MH\Svvsx
+\frac{2\SEs}{i\w}\Mvhsx\r] \nonumber \\
&&+\frac{\SEs}{\taus}\Mvhsx+\frac{\MEs}{\taus}\l(1-\frac{i\w\taus}{2}\r)\Svhsx. 
\ee
To leading order, the part in the square brackets is equivalent to $\psi_\ms$ in the
solution for $\Mvvsx$ (see Eq. \ref{phisA} and above). The leading order component of $\psi_\ms(x=0)$ is
\be
\psi_\ms\sim -\frac{\SEs^2}{4\taus^2}\frac{\lams}{\lamv} \frac{1}{i\w/2} \frac{\Mas\tauv}{i\w/2}\frac{1}{\lamv\qv}
\ee
so that we have 
\be
\Mdd\sim\Mae\tauv\frac{\SEe^2}{2\taue^2}\frac{\lame}{\lamv}\frac{1}{\sqrt{2i\w\tauv}}.
\ee

\section*{APPENDIX C. Isopotential model}
\vspace*{-3mm}
As a reference model to compare the additional effect of
spatiotemporal filtering we consider an isopotential neuron receiving
temporally filtered synaptic drive. This type of model has been
analysed previously \cite{Richardson2005} including using the upcrossing approximation
\cite{Badel2011}. The model comprises two synaptic conductances filtered
at excitatory and inhibitory time scales $\taue$ and
$\taui$. These conductances drive a voltage equation that also
includes a leak conductance. As before, it proves convenient to introduce
rate-like quantities that are conductances divided by the
membrane capacitance.
\be
\frac{dV}{dt}&=&\alpha_\ml(E_\ml-V)+H_\me(E_\me-V)+H_\mi(E_\mi-V)
\nonumber \\
\tau_\me\frac{dH_\me}{dt}&=&\alpha_\me-H_\me+\sqrt{\alpha_\me\kappa_\me}\,\xi_\me(t) \nonumber \\
\tau_\mi\frac{dH_\mi}{dt}&=&\alpha_\mi-H_\mi+\sqrt{\alpha_\mi\kappa_\mi}\,\xi_\mi(t). \label{IsoEqs}
\ee
The time-dependent quantities $\alpha_\ms(t)$ where $\ms\!=\!\me$ or $\mi$ are proportional to the presynaptic rate
whereas the $\kappa_\ms$ parameters are constant. We use a Gaussian
approximation for the synaptic drive so that $\xi_\ms(t)$ is a white-noise
process with zero mean, autocovariance
$\Ex{\xi_\ms(t_1)\xi_\ms(t_2)}=\delta(t_1-t_2)$ and it is assumed that
excitatory and inhibitory synaptic drives are uncorrelated.

Similarly to the approach used for the long-dendrite model, we
separate voltages and conductances into deterministic and zero-mean
fluctuating components $V=\Ex{V}+v$ and $H_\ms=\Ex{H_\ms}+h_\ms$. At the level of the stochastic
differential equation for voltage, we drop less significant terms that are second order in the
fluctuating components like $v h_\ms$ with the result that $v$ also
has Gaussian statistics. In terms of the quantities
$\alpha_\ms$, the deterministic equations for the isopotential neuron
are identical  to the dendritic case given in Eq. set \ref{Denddet}. The
fluctuating components, however, obey
\be
\dot{v}&=&h_\me \Ee +
h_\mi \Ei-\H v \nonumber \\
\tau_\me\dot{h}_\me&=&\sqrt{\alpha_\me\kappa_\me}\,\xi_\me-h_\me
\nonumber \\
\tau_\mi\dot{h}_\mi&=&\sqrt{\alpha_\mi\kappa_\mi}\,\xi_\mi-h_\mi
\label{IsoFluct}
\ee
where we have again the notation $\Es(t)\!=\!E_\ms\!-\!\Ex{V}$ and $\H(t)\!=\!\alpha_\ml\!+\!\Ex{H_\me}\!+\!\Ex{H_\mi}$. Note
that the difference between this isopotential reference model and the
dendritic case (Eq. set \ref{Dendfluc})
is the absence of a second spatial derivative in the equation for the
voltage and that the synaptic quantities are instead driven by
temporal Gaussian white noise not spatiotemporal Gaussian white noise.

\subsection*{Voltage-moment equations: isopotential model}
\vspace*{-3mm}
The deterministic equation set (\ref{Denddet}) provides a complete description of dynamics of the first moments
$\Ex{V}$ and $\Ex{\dot{V}}$. We now derive a set of differential
equations for the second moments of the voltage and its
derivative. First we  can solve for the variance of one of the
synaptic drives. This can be written as filter integral over the
quantity $\alpha_\ms(t)$
\be
\Ex{h_\ms^2}&=&\frac{\kappa_\ms}{\tau_\ms^2}\int_{-\infty}^{t} dt'
\alpha_\ms(t') e^{-2(t-t')/\tau_\ms} 
\ee
and because the filter is exponential, it can be rewritten in the differential form
\be
\frac{\taus}{2}\frac{d \Ex{h_\ms^2}}{dt}&=&\frac{\alpha_\ms\kappa_\me}{2\taus}-\Ex{h_\ms^2}.
 \ee
We next cross-multiply the stochastic differential equations for $v$
and $h_\ms$ by $h_\ms$ and $v$ and average to get
\be
\hspace*{-5mm}\Ex{\dot{v}h_\ms}=\Es\Ex{h_\ms^2}-\H\Ex{vh_\ms} & \mbox{and}&
\Ex{v\dot{h}_\ms }=-\Ex{vh_\ms}/\taus.
\ee
where the causality $\Ex{\xi_\ms v}\!=\!0$ has been used in the latter equation. Adding these gives the
complete derivative $\Ex{\dot{v}h_\ms}+\Ex{v\dot{h}_\ms}=\partial_t\Ex{v h_\ms}$ and so 
\be
\frac{d\Ex{v
    h_\ms}}{dt}&=&\Es\Ex{h_\ms^2}-\l(\H+\frac{1}{\taus}\r)\Ex{vh_\ms}.
\ee
We can also multiply the stochastic differential equation for $v$ by
$v$ and average to get
\be
\frac{1}{2}\frac{d\Ex{v^2}}{dt}&=&\Ee\Ex{v
  h_\me}+\Ei\Ex{v h_\mi}-\H\Ex{v^2} ~=~
\Ex{v\dot{v}}
\ee
which provide equations for both $\Ex{v^2}$ and
$\Ex{\dot{v}v}$. For the autocovariance of the rate-of-change of
voltage we multiple the differential equation for $v$ by $\dot{v}$ and average
\be
\Ex{\dot{v}^2}&=&\Ee\Ex{\dot{v}h_\me}+\Ei\Ex{\dot{v} h_\mi}-\H \Ex{v\dot{v}}. 
\ee
All together, these differential equations and subsidiary relations for the
synaptic drive and voltage provide all that is required to apply the
upcrossing method to the isopotential model.

\subsection*{Steady state: isopotential model} 
\vspace*{-3mm}
The steady state $\SV$ for the mean voltage is identical to
that given for the dendritic model; however, the variance and
variance of the rate-of-change of voltage are different. First
we note that $\Shhs=\Sas\kappa_\ms/2\taus$ and
that it is useful to use the steady-state relation
$\taus\Sdhs=\Svhs$. Then comparing the relevant equations above we have
\be
\hspace*{-5mm}\Svv&=&\frac{\SEe^2}{2}\kappa_\me\Sae\tauv
\frac{\tauv}{(\tauv+\taue)}+\frac{\SEi^2}{2}\kappa_\mi\Sai\tauv
\frac{\tauv}{(\tauv+\taui)} \label{SvvA}
\ee
which can be seen in Fig. \ref{Fig3}B (middle panel) for a case matched to the
dendritic model. For the variance of the rate-of-change of voltage we have 
\be
\hspace*{-7mm}\Sdd&=&\frac{\SEe^2}{2\taue^2}\kappa_\me\Sae\tauv
\frac{\tauv}{(\tauv+\taue)}+\frac{\SEi^2}{2\taui^2}\kappa_\mi\Sai\tauv
\frac{\tauv}{(\tauv+\taui)}
\ee
which is also illustrated in Fig. \ref{Fig3}B (lower panel). Other useful quantities are
\be
\Svhe=\frac{\SEe\Shhe}{\SH+1/\taue} &\mbox{and}& \Sdhe=\frac{\SEe\Shhe}{1+\taue\SH} 
\ee
and similarly for inhibition. 

\subsection*{Response to weak
  oscillations: isopotential model}
\vspace*{-3mm}
We again consider a weak oscillation of the excitatory drive such that
$\alpha_\me(t)=\Sae+\Mae e^{i \w t}$ and keep terms in all
calculations up to first order in $\Mae$. The deterministic,
first-order moments of the various quantities are identical to the case of the long-dendrite
considered previously. The second-order moments are different, and for the conductances we have
\be
\Mhhe=\frac{\Mae\kappa_\me}{2\taue}   \frac{1}{1+i\w\taue/2} ~\mbox{ and }~ \Mhhi=0.
\ee
The next quantites of interest are the covariances between the conductance
and voltage.
\be
\Mvhe&=&\frac{
  \SEe\Mhhe-\MV\Shhe-\MH\Svhe }{i\w +
  \SH+1/\taue} ~~\mbox{ and } \nonumber \\ \Mvhi&=&-\frac{\MV\Shhi+\MH\Svhi}{i\w +
  \SH+1/\taui} 
\ee
where $\Mhhi\!=\!0$ has been used. The oscillatory voltage variance can be
expressed in terms of these quantities
\be
\hspace*{-2mm}\Mvv\!=\!\frac{\SEe\Mvhe\!+\!\SEi\Mvhi\!-\!\MV\!\l(\Svhe\!+\!\Svhi\r)\!-\!\MH\Svv}{i\w/2\!+\!\SH}.
\ee
The covariance has the relation $\Mvd\!=\!i\w\Mvv/2$ and is therefore obtained
directly from the above. Finally, to calculate the variance of $\dot{v}$
we need 
\be
\Mdhe&=&\l(
\SEe\Mhhe-\MV\Shhe-\SH\Mvhe-\MH\Svhe
\r) \nonumber \\
\Mdhi&=&-\l(\MV\Shhi+\SH\Mvhi+\MH\Svhi
\r) 
\ee
and the same for inhibition, again noting that $\Mhhi\!=\!0$. We can then write that
\be
\hspace*{-4mm}\Mdd\!=\!\SEe\Mdhe\!+\!\SEi\Mdhi\!-\!\MV\!\l(\Sdhe\!+\!\Sdhi\r)\!-\!\SH\Mvd
\ee
where the steady-state result $\Svd\!=\!0$ has been used.

\subsection*{Low-frequency limit: isopotential model}
\vspace*{-3mm}
When $\w\!=\!0$ the $\MD$ and $\Mvd$ terms vanish as they
are time derivatives of other quantities and therefore proportional
to $\w$. It remains to calculate $\MV$, $\Mvv$ and
$\Mdd$, and when $\w\!=\!0$ these can be calculated by taking the derivatives of the steady-state values with
respect to $\Sae$. Again, it is useful to use the shorthand
$x_\me=\taue/(\tauv+\taue)$ and similarly for inhibition.
\be
\lim_{\w\to
  0}\Mvhe&=&\Mae\tauv\frac{\SEe\kappa_\me}{2\taue}x_\me\l(1-\Sae\tauv-\Sae\tauv
x_\me\r)
\nonumber \\
\lim_{\w\to
  0}\Mvhi&=&-\Mae\tauv\frac{\kappa_\mi}{2\taui}\tauv\Sai x_\mi\l(\SEe
+ \SEi x_\mi\r).
\ee
For the low frequency limit of the variance modulation we break the
response into excitatory and inhibitory components which take the form
\be
\lim_{\w\to
  0}\Mvve\!&\!=\!&\!\Mae\tauv\frac{\SEe^2\kappa_\me}{2}\frac{\tauv}{\taue}x_\me\l(1-3\Sae\tauv-\Sae\tauv
x_\me\r) \nonumber \\
\lim_{\w\to
  0}\Mvvi\!&\!=\!&\!
-\Mae\tauv\frac{\kappa_\mi}{2}\frac{\tauv}{\taui}x_\mi\!\l(2\SEe\SEi\tauv\Sai\!+\!\SEi^2\tauv\Sai\!+\!\SEi^2\Sai\tauv
x_\mi\r)\!.\nonumber
\ee
Taking a similar approach with the variance of the
rate-of-change of voltage gives
\be
\lim_{\w\to
  0}\Mdde\!&\!=\!&\!\Mae\tauv\frac{\SEe^2\kappa_\me}{2\taue^2}x_\me\l(1-2\Sae\tauv-\Sae\tauv
x_\me\r) \nonumber \\
\lim_{\w\to
  0}\Mddi\!&\!=\!&\!
-\Mae\tauv\frac{\kappa_\mi}{2\taui^2}x_\mi\l(2\SEe\SEi\Sai\tauv+\SEi^2\Sai\tauv
x_\mi\r).\nonumber
\ee

\subsection*{High-frequency asymptotics: isopotential model}
\vspace*{-3mm}
For large $\w$, the leading-order contributions can be shown to decay as $1/\w$
and comprise contributions from $\MD$, $\Mvd$ and $\Mdd$. The forms
for the first two are fairly straightforward to derive 
\be
\MD\sim\frac{\Mae\SEe}{i\w\taue}&\mbox{ and }&
\Mvd=-\frac{\Mae}{ i\w\taue
}\Svv. 
\ee
The third term is more complicated. We use
\be
\Mdhe=\SEe\Mhhe-\MH\Svhe+O\l(\frac{1}{\w^2}\r)
\ee
and similarly for $\Mdhi$ though note that
$\Mhhi$=0. Then
\be
\Mdd=\SEe\Mdhe +
\SEi\Mdhi - \SH\Mvd +O\l(\frac{1}{\w^2}\r)
\ee
where for large $\w$ we have
\be
\Mhhe\sim\frac{\Mae\kappa_\me}{i\w\taue^2} &\mbox{ and }&
\MH\sim\frac{\Mae}{i\w \taue}.
\ee
The quantities above can then be substituted into the linear response form of
the upcrossing rate, which will therefore also have a $1/\w$ behaviour at high
frequencies.

\section*{APPENDIX D. Simulations and figures}
\vspace*{-3mm}
Simulational code was written using the Julia programming language
\cite{Bezanson2017}. The code used to generate the figures is provided in the repository \textit{Gowers-Richardson-PRR-2023} at
https://github.com/mje-richardson. The simulations were implemented
using a forward Euler scheme typically with $\Delta_t\!=\!0.02$ms and $\Delta_x\!=\!20\mu$m so that
\be
&&H_\ms(x_m,t_{n+1})=H_\ms(x_m,t_n) \nonumber \\
&&+\frac{\Delta_t}{\tau_\ms}\l(\alpha_\ms(t_n)-H_\ms(x_m,t_n)+\sqrt{\alpha_\ms(t_n)\lambda_\ms}\,
\frac{\phi_{mn}^\ms}{\sqrt{\Delta_x\Delta_t}}\r) \nonumber \\
\ee
and for the voltage
\be
&&V(x_m,t_{n+1})= V(x_m,t_n)+\Delta_t\l( \alpha_\ell(E_\ell-V(x_m,t_n))\r)+ \nonumber \\
&&\Delta_t\l( H_\me(x_m,t_n)(E_\me-V(x_m,t_n))\r)+\nonumber \\
&&\Delta_t \l(H_\mi(x_m,t_n)(E_\mi-V(x_m,t_n))\r)+\nonumber \\
&&\frac{\Delta_t}{\Delta_x^2}D\l( V(x_{m-1},t_n)-2V(x_{m},t_n)-V(x_{m+1},t_n)\r) 
\ee
where $\phi_{mn}^\ms$ are independent Gaussian random numbers with zero
mean and unit variance. The system was implemented using periodic boundary conditions with
size $L\!=\!2000\mu$m being sufficiently larger than spatial correlation
lengths. Given the homogeneity of the system, statistical quantities
such as the upcrossing could be evaluated at all positions
simultaneously and averaged, thereby increasing the efficiency of
the simulations. 

For the isopotential neuron the discretisation is across time only so
the equations are
\be
&&\hspace*{-5mm}H_\ms(t_{n+1})=H_\ms(t_n)+ \nonumber \\
&&\hspace{2mm}\frac{\Delta_t}{\tau_\ms}\l(\alpha_\ms(t_n)-H_\ms(t_n) +\sqrt{\alpha_\ms(t_n)\kappa_\ms}
\,\frac{\phi_{n}^\ms}{\sqrt{\Delta_t}}\r) 
\ee
and for the voltage
\be
\hspace{-5mm}&&\hspace*{-3mm}V(t_{n+1})= V(t_n)+\Delta_t\l( \alpha_\ell(E_\ell\!-\!V(t_n))\r)+
\nonumber \\
\hspace{-5mm}&&\Delta_t\l( H_\me(t_n)(E_\me\!-\!V(t_n))\!+\!H_\mi(t_n)(E_\mi\!-\!V(t_n))\r)
\ee
where $\phi_{n}^\ms$ are again independent Gaussian random numbers with zero
mean and unit variance. 

Note that for both the dendritic and
isopotential models, the schemes above can be straightforwardly modified to
simulate the systems in the Gaussian approximation of the voltage in which terms that
are second-order in zero-mean fluctuating quantities like $v
  h_\me$ are dropped from the voltage dynamics.

\subsection*{The patterned input used in Figure 2}
\vspace*{-3mm}
The time-dependent input used in Fig. \ref{Fig2} comprised functions
$\alpha_\me(t)$ and $\alpha_\mi(t)$ lasting one second. Outside the
range $250$ to $750$ms these rates were zero. Within this range both had
constant value with $\Sae\!=\!0.00566$kHz and $\Sai\!=\!0.01100$kHz
(which would give a constant upcrossing rate of $5$Hz, anticipating
Fig. \ref{Fig3}C) with the excitatory rate $\alpha_\me(t)$ having four functions
additionally superimposed. These
functions $A(t)$ were parameterised as 
\be
A_k(t; a,t_k,\sigma,f_k)=a\exp\l(-\frac{(t-t_k)^2}{2\sigma^2}\r)\cos(2\pi f_k t)
\ee
where $a=0.03$kHz, $t_k=350,450,550,650$ms, $\sigma=20$ms and $f=0.02,0.05,0.100,0.200$kHz.

\subsection*{Illustration of steady-state properties}
\vspace*{-3mm}
Given the many components of the model, there is a broad choice of
parameter combinations that might be used to illustrate behaviour. In
the context of examining the steady-state behaviour (Fig. \ref{Fig3}B,
\ref{Fig3}C) the
choice was made to vary $\Sae$ and $\Sai$ at fixed ratio between
$\tauv$ and $\taul=1/\alpha_\ell$ to give a particular $\SV$. Given the forms
\be
\frac{1}{\tauv}&=&\alpha_\ell+\Sae+\Sai ~~\mbox{ and } \nonumber \\ 
\SV&=&\tauv(E_\ell\alpha_\ell+E_\me\Sae+E_\mi\Sai) 
\ee
we therefore have the conditions
\be
\Sae&=&\frac{(\SV-E_\mi)-(E_\ell-E_\mi)\alpha_\ell\tauv}{(E_\me-E_\mi)\tauv}
~~\mbox{ and } \nonumber \\ 
\Sai&=&\frac{(E_\me-\SV)-(E_\me-E_\ell)\alpha_\ell\tauv}{(E_\me-E_\mi)\tauv}. 
\ee
This parameter variation is used in Figs. \ref{Fig3}B and \ref{Fig3}C.

\subsection*{Matching the isopotential and dendritic models}
\vspace*{-3mm}
To provide as fair a comparison as possible between the models, we set
the parameters of the isopotential model such that the steady-state
mean voltage $\SV$, conductance state $\tauv$ and voltage variance
$\Svv$ were all matched. The mean properties of the model are
identical by design and set by $\Sae$ and $\Sai$. To match the
variance, we choose $\kappa_\me$ and $\kappa_\mi$ by comparing
Eqs. (\ref{SvvsxA}) and (\ref{SvvA}) so that 
\be
\kappa_\ms=\frac{1}{2}\frac{\lams}{\lamv}\l(\frac{\tauv+\taus}{\tauv}\r)\l(1-\sqrt{\frac{\taus}{\tauv+\taus}}\r).
\ee
Though the voltage mean and variance (Fig. \ref{Fig3}B middle panel) as well as the conductance state
parameterised by $\tauv$ are matched, it is not possible
\cite{Gowers2020} to simultaneously match the variance of the rate-of-change of
voltage (see Fig. 3B, lower panel) and so the upcrossing rates will
not be the same; this can seen in Fig. \ref{Fig3}C.


\small{

}

\end{document}